\documentclass[10pt,journal,compsoc]{IEEEtran}
\usepackage{amsmath,amsfonts}
\usepackage{algorithmic}
\usepackage{algorithm}
\usepackage{array}
\usepackage[caption=false,font=normalsize,labelfont=sf,textfont=sf]{subfig}
\usepackage{textcomp}
\usepackage{stfloats}
\usepackage{url}
\usepackage{verbatim}
\usepackage{graphicx}
\usepackage{lineno}
\usepackage{multirow}
\usepackage{float}
\usepackage{graphicx}
\usepackage{amsmath}
\usepackage{color}
\usepackage{soul}
\usepackage{booktabs}
\usepackage{threeparttable}
\usepackage{caption}
\usepackage{indentfirst}
\usepackage[T1]{fontenc}

\ifCLASSOPTIONcompsoc
  \usepackage[nocompress]{cite}
\else
  \usepackage{cite}
\fi
\ifCLASSINFOpdf
\else
\fi
\begin{document}
%
\title{Bit-balance: Model-Hardware Co-design for Accelerating NNs by Exploiting Bit-level Sparsity}
%
%
%
%

\author{Wenhao~Sun,
	Zhiwei~Zou,
	Deng~Liu,
	Wendi~Sun,
	Song~Chen,~\IEEEmembership{Member,~IEEE,}
	and~Yi~Kang

\IEEEcompsocitemizethanks{\IEEEcompsocthanksitem Wenhao Sun, Zhiwei Zou, Deng Liu, Wendi Sun, Song Chen, and Yi Kang are affiliated with the School of Microelectronics, University of Science and Technology of China, Hefei,
	Anhui, 30332 China.\protect\\
E-mail: wh1997@mail.ustc.edu.cn; songch@ustc.edu.cn
}

\thanks{ }}

\IEEEtitleabstractindextext{%
\begin{abstract}
\textbf{Bit-serial architectures can handle Neural Networks (NNs) with different weight precisions, achieving higher resource efficiency compared with bit-parallel architectures. Besides, the weights contain abundant zero bits owing to the fault tolerance of NNs, indicating that bit sparsity of NNs can be further exploited for performance improvement. However, the irregular proportion of zero bits in each weight causes imbalanced workloads in the Processing Element (PE) array, which degrades performance or induces overhead for sparse processing. Thus, this paper proposed a bit-sparsity quantization method to maintain the bit sparsity ratio of each weight to no more than a certain value for balancing workloads, with little accuracy loss. Then, we co-designed a sparse bit-serial architecture, called Bit-balance, to improve overall performance, supporting weight-bit sparsity and adaptive bitwidth computation. The whole design was implemented with 65nm technology at 1 GHz and performs at \boldmath{$326$-, $30$-, $56$-, and $218$-}frame/s for AlexNet, VGG-16, ResNet-50, and GoogleNet respectively. Compared with sparse bit-serial accelerator, Bitlet, Bit-balance achieves \boldmath{$1.8\times$\textasciitilde$2.7\times$} energy efficiency (frame/J) and \boldmath{$2.1\times$\textasciitilde$3.7\times$} resource efficiency (frame/mm$^2$)}.
\end{abstract}

\begin{IEEEkeywords}
hardware accelerator, bit sparsity, quantization, neural network.
\end{IEEEkeywords}}

\maketitle

\IEEEdisplaynontitleabstractindextext

%
\IEEEpeerreviewmaketitle

\IEEEraisesectionheading{\section{Introduction}\label{sec:introduction}}
Nowadays, NNs have been widely applied in numerous domains, such as image recognition\cite{10.1145/3065386}, \cite{simonyan2015deep},\cite{7298594},\cite{7780459},\cite{7780460} speech recognition\cite{7178838}, object detection\cite{7780460}, and computer vision\cite{10.1145/2647868.2654889}. Structurally, it mainly consists of convolutional (CONV) layers and fully connected (FC) layers; the former conduct convolution, and the latter conduct matrix-vector multiplication (GEMV), both of which include massive multiplication-accumulation (MAC) operations.  As higher precision and complication demands arise, the workloads of network computing continue to increase. Therefore, researchers have been striving for neural network hardware accelerators to catch up with software development.	\\ 
\indent\setlength{\parindent}{1em}Although NNs bring intensive computations, there exist monotonous operations; thus, many hardware architects attempt to reduce runtime by improving computation parallelism. The accelerators can be classified as four categories, bit-parallel, sparse bit-parallel, bit-serial, sparse bit-serial. DaNN\cite{7011421}, Eyeriss\cite{7738524}, TPU\cite{8192463}, Tianjic\cite{8998338}, and etc are typical bit-parallel accelerators, which primarily actualizes acceleration through improving parallelism with massive fixed-point computing units. Additionally, since the input feature maps (IFMs) and weights contain abundant zero elements owing to the fault tolerance of NNs, which can be exploited to accelerate NNs by skipping zero-elements computing. Therefore, lots of sparse bit-parallel accelerators emerge, such as EIE\cite{7551397}, Cambricon-X\cite{7783723}, Cnvlutin\cite{7551378}, SCNN\cite{8192478}, etc, further improving performance and energy efficiency further. However, these works primarily are designed for fixed-point computing, without consideration of varying weight/IFM-precision for different NNs, causing inefficient resource utilization and energy efficiency. \\ 
\indent\setlength{\parindent}{1em}To adapt to varying precisions of different NNs, researchers attempt to design accelerators based on bit-serial computing.  Stripe\cite{7529197} presented a hardware accelerator that relied on bit-serial computing units and quantizes IFMs to the required precision, $p$, in each layer, which can provide $16/p\times$ speedup compared with 16-bit fixed-point baseline. Loom\cite{8465915} reduced the number of both weights and IFMs bits for bit-serial acceleration. Bit-Fusion\cite{8416871} proposed a bit-flexible accelerator to match the bitwidth of each individual multiply-add operand, maximizing the resource efficiency for operations of different bitwidths. UNPN\cite{8481682} designed a Lookup table(LUT)-based bit-serial PE, achieving 23\%~54\%  energy consumption reduction compared with conventional fix-point MAC units. These accelerators leverage bit-serial computing to improve performance and energy efficiency. However, they didn't exploit bit sparsity, indicating there are still room for performance improvement. \\ 
\indent\setlength{\parindent}{1em}Owing to the bit sparsity of NNs, Some researchers designed sparse bit-serial accelerators to further improve performance. Pragmatic\cite{8686550} skips the computing at the common zero-bit position of involved IFM elements. But the zero-bits distribution of each IFM bit sequence is irregular, resulting in only a small propotion of truncated zero-bits. Laconic\cite{8980351} tears MAC operation into IFMs and weights bit-serial computing and accelerates by exploiting  both IFMs and weights sparsity. However, the distribution of zero bit in IFMs and weights can be irregular, which causes imbalanced workload in the PE array and degrades performance. Bit-Tactical\cite{10.1145/3297858.3304041} applied a weight skipping module and an IFM-bit selection module to optimize the layout of irregular sparse weights through front-end scheduling, thus further avoiding the calculation of zero weight elements and IFM bits. However, the independent supply of IFMs between PEs makes its data cache resources far greater than the area of computing resources, resulting in low resource utilization. Bitlet\cite{10.1145/3466752.3480123} proposes the bit interleaving philosophy to maximally exploit bit-level sparsity, which enforces acceleration by decreasing the number of weights involved in computing. However, the logic corresponding to bit interleaving occupies over 40\% area of the entire PE module. Besides, the computing units of high precision can be idle during low precision operation, causing inefficient resource efficiency. Though these architectures exploit bit-sparsity for acceleration, they still suffer from imbalanced workload or huge overhead for implementation of sparse bit-serial processing. \\ 
\indent\setlength{\parindent}{1em} This paper aims to balance sparse bit-serial workloads in the PE array while lowering the overhead of sparsity processing. A model-hardware co-design of the sparse bit-serial accelerator for NNs is proposed to improve system performance and energy efficiency. Our main contributions are as follows: \\ 
\indent\setlength{\parindent}{1em}1). We proposed a systolic-array-based accelerator, called Bit-balance, supporting sparse bit-serial processing of weights and adaptive bitwidth computing, achieving high performance without extra preprocessing module by processing the sparse weights of the encoding format directly.\\ 
\indent\setlength{\parindent}{1em}2). We co-designed a bit-sparsity quantization method to keep the sparsity ratio of each weight at a certain value with little accuracy loss, which can balance PE loads across the PE array and significantly improve performance. \\ 
\indent\setlength{\parindent}{1em}Compared with bit-serial accelerator, Stripe\cite{7529197}, Bit-balance exploit weight bit sparsity for acceleration and achieved $4\times$\textasciitilde$7.1\times$ speedup and $3\times$\textasciitilde$5.6\times$ energy efficiency. Besides, Compared with sparse bit-serial accelerator, Laconic\cite{8980351}, we balanced the workload in the PE array and achieved $2.2\times$\textasciitilde$4.3\times$ speedup and $2.7\times$\textasciitilde $5.4\times$ energy efficiency. Further, compared with Bitlet\cite{10.1145/3466752.3480123}, we lowered the overhead of sparse processing significantly, achieving $1.8\times$\textasciitilde $2.7\times$  energy efficiency and $2.1\times$\textasciitilde $3.8\times$ resource efficiency. \\

\section{Preliminary}
Since NNs can be applied to various domains, the precisions of weights are diverse. Besides, weights of NNs contain many abundant zero bits due to the strong fault tolerance. Thus, there is great potential for architectures to gain more benefits based on bit-serial computing. Although bitwidth quantization  can decrease the bit-serial computing cycles directly, only limited bitwidth can be cut for maintaining accuracy. Exploiting the bit sparsity of weights can further accelerate bit-serial computing, but the PE array suffers from imbalanced workloads. Those PE processing weights with a small number of non-zero bits (NNZB) must wait for those with large NNZB, which will degrade the performance and PE utilization. Therefore, the key challenge is to balance the sparse bit-serial computing workloads across the PE array.

\subsection{Computing Process of NN}

\begin{figure}[H]
	\centering
	\includegraphics[width=0.8\linewidth]{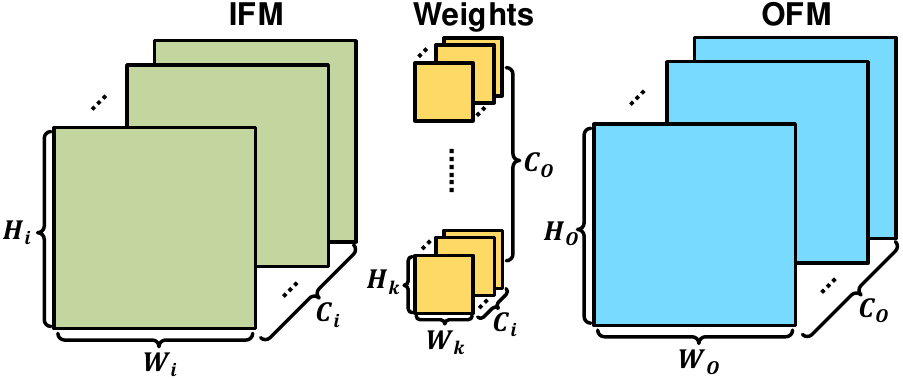}
	\caption{ Convolution process. }
	\label{fig1}
\end{figure}

The computing process of NNs is shown in Fig.\ref{fig1}. A CONV layer receives $C_i$ input channels (ICs) of IFMs with a size of $H_i \times W_i$. Each OFM of $C_o$ output channels (OCs), with a size of $H_o \times W_o$, is generated by sliding a $H_k \times W_k$ kernel window on one IFM and then accumulating across the temporary results of $C_i$ ICs. Finally, $C_o$ OFMs are calculated by convolving $C_o$ groups of kernels with shared IFMs. For FC layers, the size of IFMs and OFMs are both $1\times1$. Set IFMs as matrix $I$[$C_i$,$H_{i}$,$W_{i}$], weights as $W$[$C_o$,$C_i$,$H_{k}$,$W_{k}$] and OFMs as $O$[$C_o$,$H_o$,$W_o$], and the process can be described as Equ.\ref{equ1}.	
\begin{equation}
	\label{equ1}
	\begin{aligned}
		O[o,x,y] = \sum_{i=0}^{C_i} \sum_{a=0}^{H_k} \sum_{b=0}^{H_k} I[i,x+a,y+b] \times W[o,i,a,b] \\
		(0 \leq o < C_o, 0 \leq x < H_{o}, 0 \leq y < W_{o})
	\end{aligned}
\end{equation}

Referring to Equ.\ref{equ1}, the basic operations of NN processing are MACs. Assuming the bitwidth of weights is $N$, the MAC of an IFM, $I_0$, and a weight, $W_0$, can be decomposed to shift and add operations, as shown in Equ.\ref{equ2}, which is flexible for varying bitwidth.	 

\begin{equation}
	\label{equ2}
	\begin{aligned}
		I_0\times W_0 = \sum_{i=0}^{N} I_0 \times W_0[i] \times 2^i 
	\end{aligned}
\end{equation}

\subsection{Challenges of Sparse bit-serial computing}
\begin{figure}[H]
	\centering
	\includegraphics[width=0.99\linewidth]{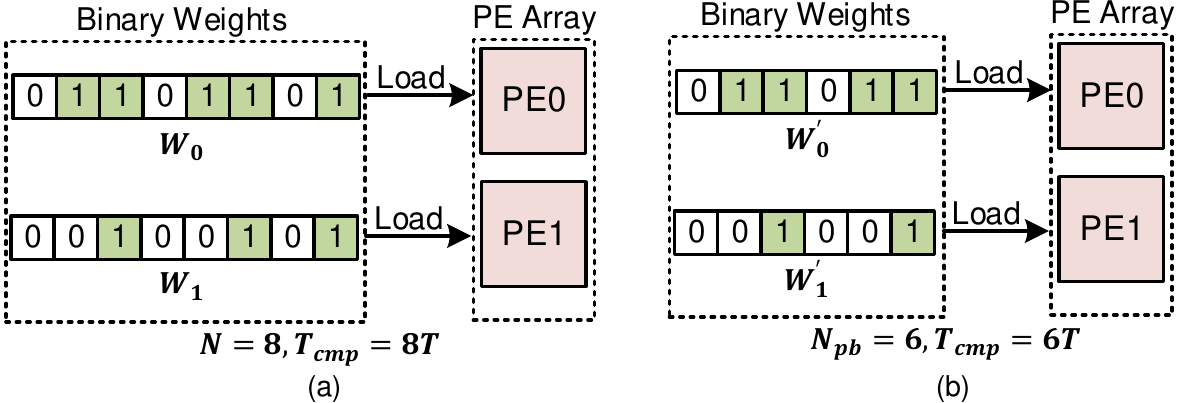}
	\caption{Bit-serial computing with original weights and quantized weights in the PE array. (a)Original weights. (b)Quantized weights.}
	\label{fig2}
\end{figure}

\begin{figure}[H]
	\centering
	\includegraphics[width=0.99\linewidth]{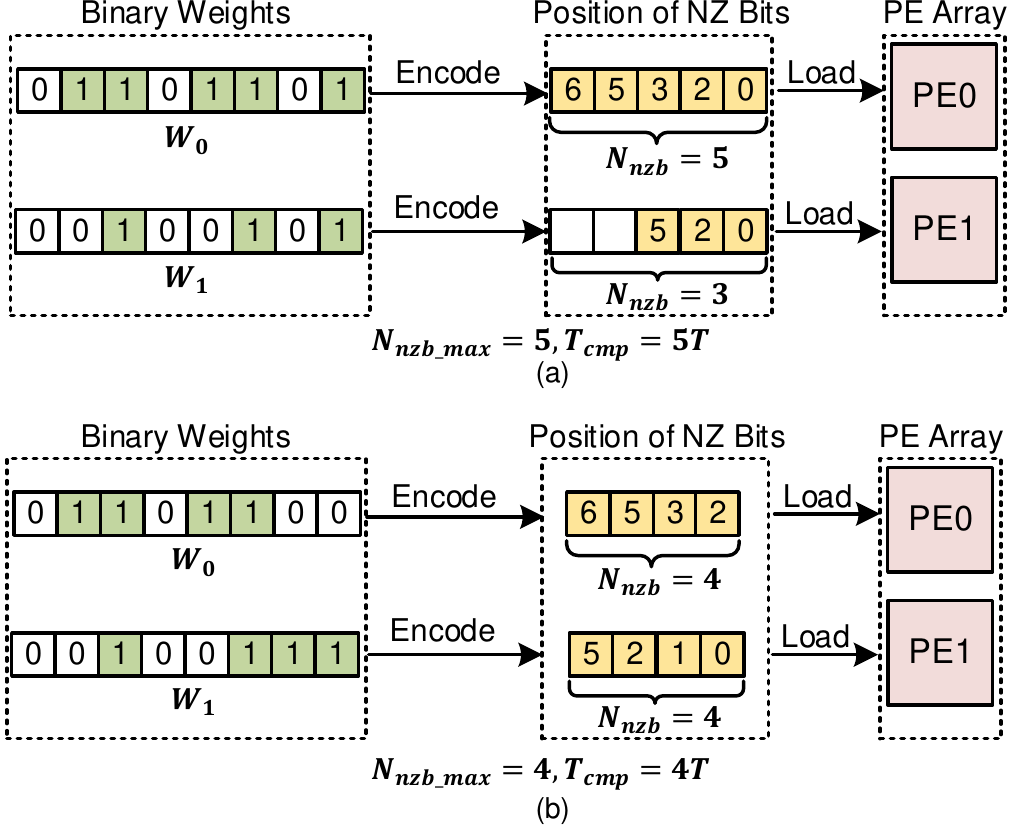}
	\caption{Sparse bit-serial computing with imbalanced and balanced workloads of non-zero (NZ) weight bits in the PE array. (a)Imbalanced workloads. (b)Balanced workloads.}
	\label{fig3}
\end{figure}

To accelerate bit-serial processing, quantizing the weight to lower bitwidth is the most straightforward method. If the weight is quantized from $N$-bit to $N_{pb}$-bit, we can achieve $N/N_{pb}\times$ speedup. An example of bit-serial computing with quantized weight is shown in Fig.\ref{fig2}, achieving $1.3\times$ performance improvement. However, assuming the value of $N$-bit weight is $2^{N}$, it can be decreased to $2^{N_{pb}}$ when the bitwidth is tailored to $N_{pb}$. To maintain the accuracy, the reduction of bitwidth can be limited, which constrains the performance improvement. \\
\indent\setlength{\parindent}{1em} To further improve performance, the bit sparsity of weights can be exploited for acceleration by skipping zero-bits computing. Assuming the NNZB in each weight is $N_{nzb}$ with the largest value being $N_{nzb\_max}$, weights are compressed and loaded in PE array, with $N_{nzb\_max}$ computing cycles. An example of bit-serial computing by exploiting sparsity is shown in Fig.\ref{fig3}(a), achieving $1.6\times$ performance. However, since the sparsity ratio of weights is randomly distributed, the workloads across the PE array is imbalanced. The PEs with small NNZB must wait for those with large NNZB to finish. Thus, $PE_1$ will be idle for $2$ cycles, indicating inefficient PE utilization. Moreover, once the $N_{nzb\_max}$ is equal to original weight bitwidth, there can be no performance improvement.

\indent\setlength{\parindent}{1em}Although there are some previous works of bit-serial computing acceleration, such as Stripe\cite{7529197}, Laconic\cite{8980351} and Bitlet\cite{10.1145/3466752.3480123}, they all suffer from performance degradation or inefficient resource efficiency owing to imbalanced workloads. Thus, in this work, we aim to balance the workloads of sparse weight bits in the PE array.

\section{Methodology}

To deal with the imbalanced workloads in the PE array, we quantize the weights based on bit sparsity and maintain the NNZB in each weight to no more than a certain value, achieving significant improvement of PE utilization and performance. To fit with bit-serial computing, we store the weights in encoding format, which can be used directly for shift-add operation. 

\subsection{Bit-Sparsity Quantization}

\begin{figure}[H]
	\centering
	\includegraphics[width=0.8\linewidth]{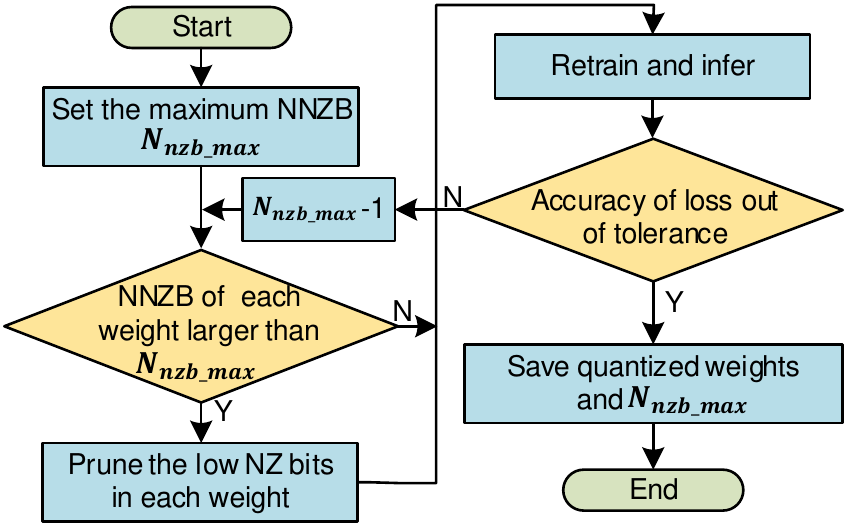}
	\caption{The flow of bit-sparsity quantization.}
	\label{fig4}
\end{figure}

\begin{figure}[H]
	\centering
	\includegraphics[width=0.93\linewidth]{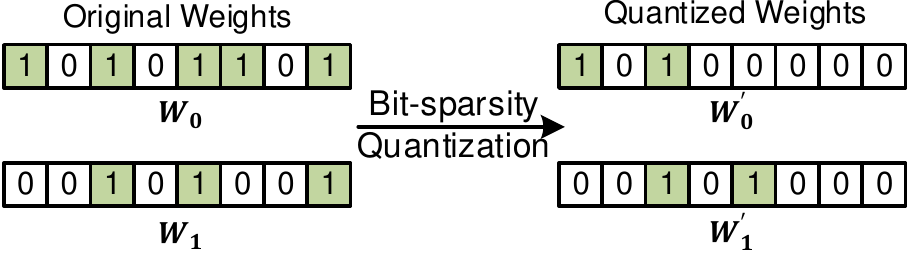}
	\caption{Example of bit-sparsity quantization. }
	\label{fig5}
\end{figure}

Since NNs have the nature of strong fault tolerance, the weights allow redundant bits. However, quantizing weights to lower bitwidths directly will significantly decrease the numeric range of weight value and exploiting the randomly distributed bit sparsity of weights will suffer from imbalance workloads. Thus, to maintain the numeric range of weight value and balance computing loads across the PE array, we proposed a bit-sparsity quantization method, which set several weight bits as zero and maintain the maximum NNZB, $N_{nzb\_max}$, in each weight to no more than a certain value. Assuming the bitwidth of original weight is $N$, its numeric range can be calculated as $\sum_{i=0}^{N_{nzb\_max}}\binom{N}{i}$.  Compared with quantizing the weights to $N_{pb}$-bit directly, the numeric range of weight value in our method is more abundant, as shown in Tab.\ref{tab:tab1}. The case of $N_{nzb\_max} = 3$ in bit-sparsity quantization can be competitive with $N_{pb}=9$ in bitwidth quantization. \\
\begin{table*}\centering
	\caption{Numeric Range Comparison of Weight Values}
	\label{tab:tab1}
	\begin{tabular}{|c|c|c|c|c|c|c|c|c|c|c|c|c|}
		\hline
		\multirow{2}{*}{\begin{tabular}[c]{@{}c@{}}Bitwidth  \\Quantization \end{tabular}} & $N_{pb}$           & \textbf{13}                          & \textbf{12}                          & \textbf{11}                          & \textbf{10}                          & \textbf{9}                           & \textbf{8}      & \textbf{7}      & \textbf{6}      & \textbf{5}      & \textbf{4}      & \textbf{3}      \\ \cline{2-13} 
		& $2^{N_{pb}}$            &         8192                    &        4096                     &        2048                     &         1024                    &        512                     & 256    & 128    & 64     & 32     & 16     & 8      \\ \hline
		\multirow{2}{*}{\begin{tabular}[c]{@{}c@{}}Bit-sparisty \\ Quantization\end{tabular}}      & ($N_{nzb\_max}$,$N$) & \multicolumn{1}{c|}{(\textbf{13},16)} & \multicolumn{1}{c|}{(\textbf{12},16)} & \multicolumn{1}{c|}{(\textbf{11},16)} & \multicolumn{1}{c|}{(\textbf{10},16)} & \multicolumn{1}{c|}{(\textbf{9},16)} & (\textbf{8},16) & (\textbf{7},16) & (\textbf{6},16) & (\textbf{5},16) & (\textbf{4},16) & (\textbf{3},16) \\ \cline{2-13} 
		& $\sum_{i=0}^{N_{nzb\_max}}\binom{N}{i}$           & 65339                       & \multicolumn{1}{c|}{64839}  & \multicolumn{1}{c|}{63019}  & \multicolumn{1}{c|}{58651}   & \multicolumn{1}{c|}{50643}   & 39203    & 26333  & 14893  & 6885   & 2517   & 697    \\ \hline
	\end{tabular}
\end{table*}
\indent\setlength{\parindent}{1em}Compared with exploiting the randomly distributed bit sparsity of weights, the performance based on our quantization method is higher owing to the more balanced workload of sparse weight bits across the PE array. An example of the computing flows toward bit-sparsity quantization is shown in Fig.\ref{fig2}(b). The NNZB in $W_0$ and $W_1$ are both four, eliminating the idle PEs. Therefore, we obtained the total computing time of $4T_w$ with $1.25\times$ speedup compared with the imbalanced workload case.\\
\indent\setlength{\parindent}{1em}The specific flow of bit-sparsity quantization is shown in Fig.\ref{fig4}. First, we set an initial maximum NNZB of weights, $N_{nzb\_max}$. Then, for weights with NNZB larger than $N_{nzb\_max}$, we set the less significant none-zero bits as zero and maintain the total NNZB to $N_{nzb\_max}$. Next, the previously quantized weights were retrained and testified if the accuracy dropped out of boundary. If not, we continued to decrease $N_{nzb\_max}$ and train the weights; otherwise, we saved the final quantized weights and recorded $N_{nzb\_max}$. Fig.\ref{fig5} shows a quantization example of two 8-bit weights, $W_0$ and $W_1$. We set the NNZB in each weight to less than $4$. Although the distribution of zero bits is irregular, the workloads of each weight are balanced, thereby achieving $8/4 = 2\times$ speedup.   

\subsection{Sparse Bit-serial Processing}
\begin{figure}[H]
	\centering
	\includegraphics[width=0.4\linewidth]{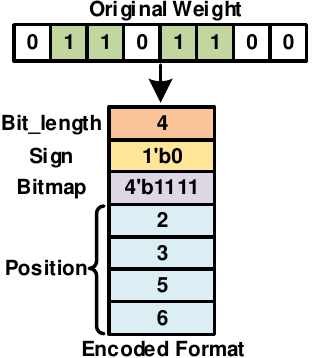}
	\caption{Encoding format of weights.}
	\label{fig6}
\end{figure}

To reduce the overhead of sparsity processing, the weights are encoded with length, sign, bit position, and bitmap, as shown in Fig.\ref{fig6}. The length, $N_{nzb\_max}$, represents the maximum NNZB of all weights, which only needs to be stored once for all weights in the current layer. The sign, $W_s$, indicates the positivity or negativity of one weight. The bit position, $W_{p}$, determines how many bits IFM should shift. The bitmap, $W_b$, indicates whether the bit position is valid or not since the NNZB of some weights is smaller than $N_{nzb\_max}$.

\begin{figure}[H]
	\centering
	\includegraphics[width=0.9\linewidth]{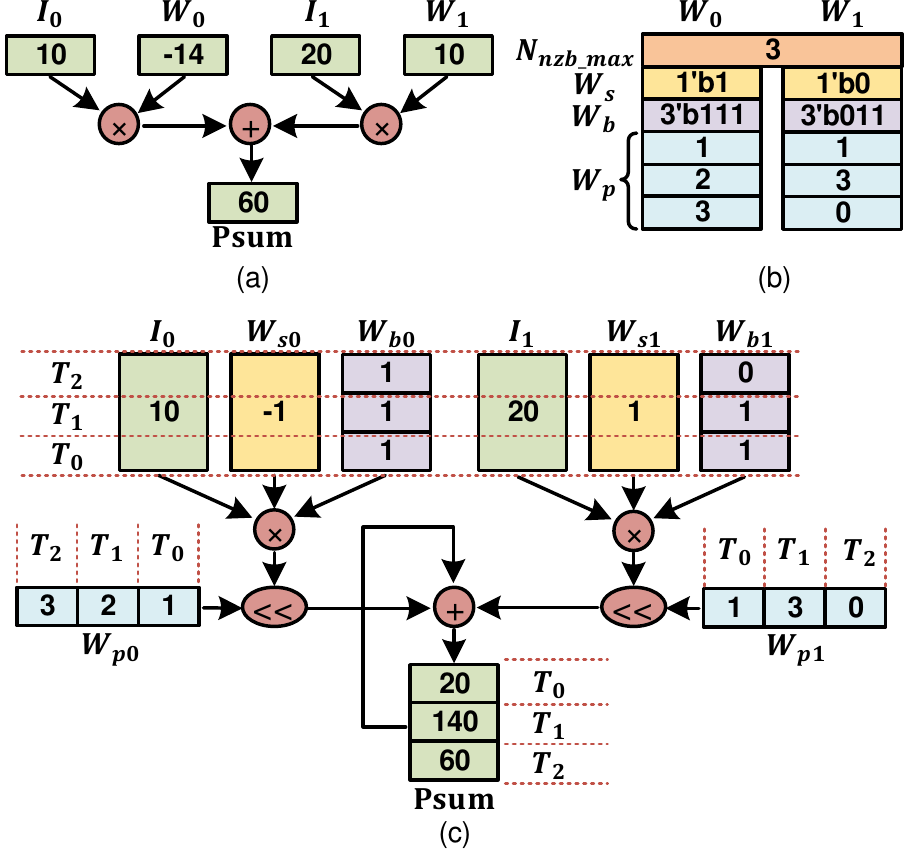}
	\caption{An example of sparse bit-serial computing. (a)Original bit-parallel computing. (b)Encoded Weights. (c)Process of sparse bit-serial computing.}
	\label{fig7}
\end{figure}
Fig.\ref{fig7} shows an example of sparse bit-serial computing. Fig.\ref{fig7}(a) shows the original MAC operation, $Psum = I_{0}*W_{0}+I_{1}*W_{1}$. Fig.\ref{fig7}(b) shows the encoding format of two weights, with the maximum NNZB, $N_{nzb\_max}=3$. The computing flow is shown in Fig.\ref{fig7}(c). At $T_0$, we fetch $W_{p0} = 1$, $W_{p1} = 1$, and then $I_{0}$ and $I_{1}$ both shift left 1-bit, calculating $Psum = 20$. Similarly, at $T_1$, the Psum is calculated as $140$. Since the last bitmap of $W_{b1}$ is zero, the operation of $W_{p1}$ is invalid and the final result is $60$. The entire process takes $3$ cycles, which is $2.67\times$ faster than $8$ cycles of the original process.

\section{Bit-balance Architecture}

To achieve high speedup and energy efficiency at low hardware cost, we chose the systolic array\cite{1653825} as the mainstay for high bandwidth saving and simplified inter-PE connection. To further reduce the overhead of sparse processing, the weights were encoded at software level, without introducing the corresponding preprocessing module. The control of sparse bit-serial computing is mastered by top controller based on the maximum NNZB. Thus, the PEs in our architecture only requires the shift-add units. The design of the overall architecture aims to balance the workloads of sparse weight bits at a low hardware cost, collaborating with bit-sparsity quantization.
\begin{figure}[H]
	\centering
	\includegraphics[width=0.81\linewidth]{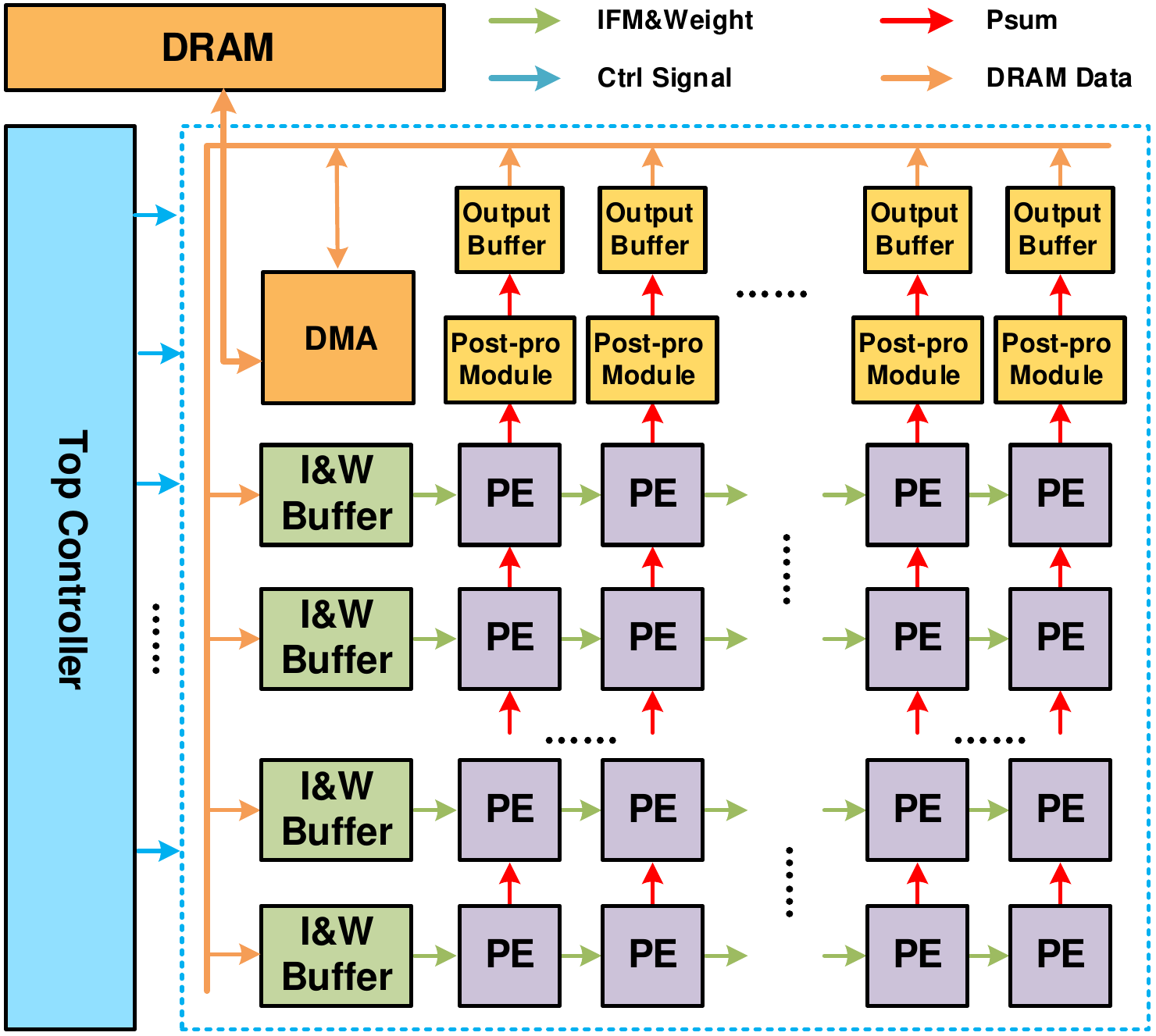}
	\caption{Top-level architecture.}
	\label{fig8}
\end{figure}

\subsection{Overview}

As shown in Fig.\ref{fig8}, the top-level architecture mainly consists of input-weight buffers (I\&W buffer), a systolic-based PE array, post-processing modules (Post-pro module), output buffers, a top controller, and a DMA interface. Initially, the weights in encoding format and control parameters were generated at software level. These data and input images were directly loaded into DRAMs. Next, the IFMs and weights were fetched by I\&W buffers and then transferred to the PE array. After massive bit-serial operations by PEs, post-pro modules take in the outputs to perform ReLU and pooling for final OFMs, which were buffered in the output buffers temporarily. Finally, the OFMs of the current layer were stored to DRAMs for the next layer computing. We executed layer-wise, and the OFMs of the last layer were classified to output the final results. The top controller mastered the entire computing flow and the on/off-chip data were transported through the DMA interface.

\subsection{Systolic-based PE Array}
\begin{figure}[H]
	\centering
	\includegraphics[width=0.7\linewidth]{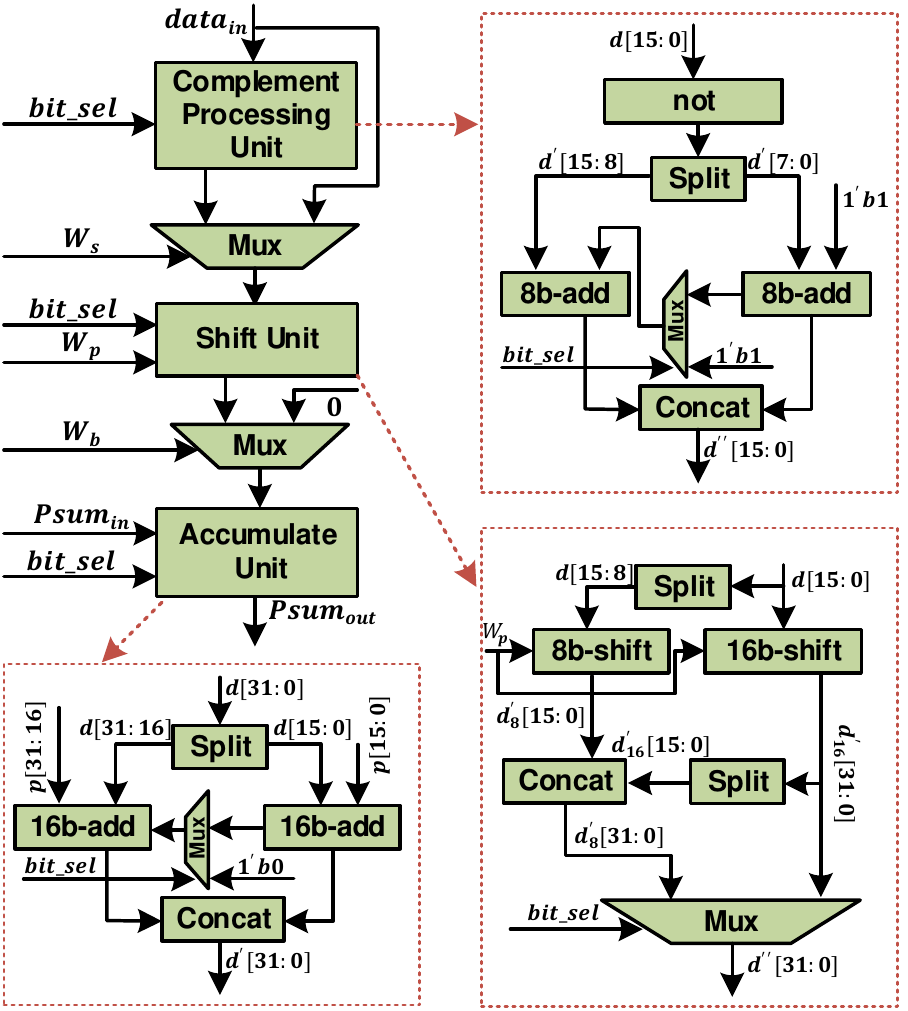}
	\caption{PE unit.}
	\label{fig9}
\end{figure}

The systolic-based PE array, in charge of bit-serial computing, is composed of $N_{PE} \times N_{PE}$ PEs, among which PEs of the same row share IFMs of one IC, and PEs of the same column process OFMs of one OC. Each PE contains a complement processing unit, a shift unit, and an accumulation unit, as shown in Fig.\ref{fig9}. First, the sign of weight, $W_s$, chose the complement or the original value of the IFM. Then, the temporary result was shifted left based on the weight-bit position, $W_{p}$. Finally, the shifted result was accumulated with adjacent PE. To reduce power consumption, the complement processing unit and shift unit are gated when the weight bitmap, $W_{b}$, is zero. Moreover, to adapt to different bitwidths of IFMs and weights, we merged the 8-bit operation with 16-bit operation for higher resource utilization. The 16-bit IFMs with the corresponding 32-bit temporary results can be divided as two pairs of 8-bit IFMs and 16-bit temporary results, achieving $2\times$ peak throughput in 8-bit operation compared with the 16-bit.

\subsection{On-chip Buffer}
To reduce energy consumption of the on/off-chip data movement, we need to buffer IFMs, weights, and OFMs for data reusing. Thus, the on-chip buffers mainly consist of I\&W buffers and output buffers. To avoid performance bottleneck of data loading, we utilized each group of two memories as Ping-Pong buffers, one for DRAM access and the other for PE transporting, which is similar with other systolic-array based accelerators\cite{Sense}. To support adaptive bitwidth computing, the bitwidths of I\&W buffers and output buffers are both 16-bit, which can store one 16-bit IFM/OFM or two 8-bit IFMs/OFMs per address. For weights in encoding format, we store 16 weight signs, 16 weight bitmaps, 4 weight positions of 16-bit precision, or 5 weight positions of 8-bit precision per address.

\section{Dataflow of Sparse Bit-serial Processing}

To reduce the energy consumption of DRAM access, we applied a dataflow to enhance data reusing with limited on-chip buffer resources. Since Bit-balance is a systolic-array-based accelerator, we merge the sparse bit-serial computing with systolic dataflow\cite{Sense}. Furthermore, a mapping algorithm was designed to map various networks on our architecture, based on provided network parameters.

\subsection{Data Partition}
For limited on-chip buffer resources and PE array size, we need to partition IFMs and weights into independent sub-tiles in terms of edge size and input/output channel, referring to the tiling method\cite{Sense}. Considering the resources of Psum storage, the size of IFM sub-tiles is set no larger than $W_{IS} \times H_{IS}$, and the sizes of IFM tiles on the width and height dimension are $T_{WI}$ and $T_{HI}$, respectively. For limited PE array size, we only process $N_{PE}$ input and output channels concurrently, with tiling numbers of $T_{IC}$ and $T_{OC}$. The specific tiling parameters are shown in Tab.\ref{tab:tab2}. After data partition, each tile of IFMs and weights can be processed with sparse bit-serial computing independently. 

\begin{table}[H]\centering
	\caption{Parameters of Data Partition}
	\label{tab:tab2}
	\begin{tabular}{|c|c|c|}
		\hline
		Parameter & Explanation                & Formula \\ \hline
		$N_{PE}$       & PE array size              & -       \\ \hline
		$N_{IC}$       & \# of input channel        & -       \\ \hline
		$N_{OC}$       & \# of output channel       & -       \\ \hline
		$W_{K}$        & Width of a kernel              & -       \\ \hline
		$H_{K}$        & Height of a kernel              & -       \\ \hline
		$W_{I}$        & Width of a IFM              & -       \\ \hline
		$H_{I}$        & Height of a IFM              & -       \\ \hline
		$W_{IS}$       & Width of a IFM tile       & -       \\ \hline
		$H_{IS}$       & Height of a IFM tile       & -       \\ \hline
		$T_{IC}$        & Tiling \# of input channel & $N_{IC}$/$N_{PE}$ \\ \hline
		$T_{OC}$       & Tiling \# of input channel & $N_{OC}$/$N_{PE}$ \\ \hline
		$T_{WI}$   & Tiling \# of IFM width    & $W_{I}$/$W_{IS}$  \\ \hline
		$T_{HI}$   & Tiling \# of IFM height    & $H_{I}$/$H_{IS}$  \\ \hline
	\end{tabular}
\end{table}
\subsection{Mapping Algorithm }

Considering dataflow and partition of IFM and weight, to map various networks on Bit-balancing, we designed a network mapping algorithm. The computing flow description is shown in Tab.\ref{tab:tab3}. The 1st and 4th rows decide the dataflow according to Adaptive Dataflow Configuration\cite{Sense}. For "reuse IFM first (RIF)", we finish the OFM computing of all OCs for one output tile before switching to the next output tile; otherwise, for "reuse weight first (RWF)", we finish computing of all output tiles for one OC before switching to the next OC. The 2nd and 3rd rows describe the partition of IFM, with a row number of $T_{ofm\_row}$ and column number of $T_{ofm\_col}$. The 5th row drives the accumulation of input channels to obtain the final Psums. The 6th and 7th rows represent the element number of one tile of IFM and one kernel. The 8th row indicates the maximum NNZB of the weights, where $N_{nzb\_max}$ is loaded as a parameter because the weights from training are fixed. The 9th row indicates that the PE is performing the sparse bit-serial computing with IFM and the weight bit position.

\begin{table}[H]\centering
	\caption{Computing Flow}
	\label{tab:tab3}
	\begin{tabular}{lll}
		\toprule
		\textbf{Input}: parameters of architecture configuration; IFMs; weights;\\
		\textbf{Output}: OFMs \\
		\midrule
		Noting: When RIF, $T_{OC\_RWF} = 1, T_{OC\_RIF} = Toc;$ otherwise,\\
		$T_{OC\_RWF} = Toc, T_{OC\_RIF} = 1$\\
		\textbf{1}\hspace{0.2cm}$for(a=0;a<T_{OC\_RWF};a=a+1)$	\\
		\textbf{2}\hspace{0.4cm}$ for(b=0;b<T_{WI};b=b+1)$	\\
		\textbf{3}\hspace{0.6cm}$  for(c=0;c<T_{HI};c=c+1)$	\\
		\textbf{4}\hspace{0.8cm}$   for(d=0;f<T_{OC\_RIF};d=d+1)	$\\
		\textbf{5}\hspace{1cm}$      for(e=0;e<T_{IC};e=e+1)$ \\
		\textbf{6}\hspace{1.2cm}$     for(f=0;f<W_{K}\times H_{K};f=f+1)$	\\
		\textbf{7}\hspace{1.4cm}$ 	    for(g=0;g<W_{K}\times H_{K};g=g+1)$	\\
		\textbf{8}\hspace{1.6cm}$ 	    for(h=0;h<N_{nzb\_max};h=h+1)$	\\
		\textbf{9}\hspace{1.8cm}$	     Psum+= I_{NZ}\times2^{W_{p}}$\\
		
		\bottomrule
	\end{tabular}
\end{table}

\section{Experiments}

\subsection{Implementation}

Bit-balance Implementation: To measure the area, power, and critical path delay, we implemented Bit-balance in Verilog and conducted a functional simulation on Cadence Xcelium 2019.2. Then, the Verilog code was synthesized with Synopsys Design Compiler (DC). The on-chip SRAM buffers are generated by Memory Compiler. The on-chip power is calculated by Synopsys Prime Suite 2020.09 with the Switching Activity Interface Format(SAIF) file generated from testbench. The reports of resource utilization and power breakdown are shown in Tab.\ref{tab:tab4}.\\
\indent\setlength{\parindent}{1em}The size of the PE array is $32\times32$, achieving a peak throughput of $1024GOP/s$ and $2048GOP/s$ for 16b and 8b shift-add operations respectively. The IFMs and weights are both encoded in 16 bits or 8bit, and their Psum are correspondingly truncated to 32 bits and 16 bits respectively. Since the PEs are implemented with shift-add units, the area of computing core (CC), containing PE array, post processing module, and controller, is only 2.91{mm$^{2}$}. To balance the data reuse efficiency and resource overhead, we set the IFM tiling unit as $8\times8$. Each I\&W buffer contains two $1K\times16b$ and two $256\times16b$ single-port static random-access memories (SRAMs); each output buffer and post processing module contain two $64\times16b$ single-port Register Files (RFs) and one $64\times16b$ dual-port RF, respectively. Thus, the total on-chip memories are 176KB. The power consumption mainly comes from the computing core, taking up 85\% power of full chip. \\

\begin{table}[H]\centering
	\caption{Resource and Power Proportion\iffalse of Each Module\fi}
	\label{tab:tab4}
	\begin{tabular}{|c|c|c|c|}
		\hline
		Module          & \begin{tabular}[c]{@{}c@{}}Area\\ (mm2)\end{tabular} & \begin{tabular}[c]{@{}c@{}}16b-Power\\ (mW)\end{tabular} & \begin{tabular}[c]{@{}c@{}}8b-Power\\ (mW)\end{tabular} \\ \hline
		I\&W Buffer     & 1.81(36.3\%)                                         & 90(11\%)                                                 & 86(10\%)                                                \\ \hline
		PE array        & 2.34(46.9\%)                                         & 582(71\%)                                                & 609(71\%)                                               \\ \hline
		Post-pro Buffer & 0.49(9.8\%)                                          & 82(10\%)                                                 & 95(11\%)                                                \\ \hline
		Output buffer   & 0.27(5.4\%)                                          & 41(5\%)                                                  & 42(5\%)                                                 \\ \hline
		Controller      & 0.08(1.6\%)                                          & 25(3\%)                                                  & 25(3\%)                                                 \\ \hline
		Computing core  & 2.91(58.3\%)                                         & 689(84\%)                                                & 729(85\%)                                               \\ \hline
		Total           & 4.99                                                 & 820                                                      & 857                                                     \\ \hline
	\end{tabular}
\end{table}

To verify the superiority of Bit-balance, we set some baseline accelerators as shown in Tab.\ref{tab:tab5}. The brief descriptions of each accelerator are presented as below:\\		
\indent\setlength{\parindent}{1em}1) Eyeriss\cite{7738524} is a typical bit-parallel accelerator based on the 16-bit fixed-point MAC, without considering the sparsity of IFMs or weights. By comparison, Bit-balance achieves better performance through exploiting weight bit sparsity and higher energy efficiency by replacing MAC units with shift-add units.\newline
\indent\setlength{\parindent}{1em}2) Cambricon-X\cite{7783723} accelerates NNs by skipping zero-weight elements computing. However, the whole architecture is designed for accelerating NNs of 16-bit precision, resulting in inefficient resource utilization in lower-bit operation. By comparison, Bit-Balance achieves higher resource efficiency through bit-serial computing.\\
\indent\setlength{\parindent}{1em}3) Stripe\cite{7529197} allows per-layer selection of precision instead of fixed-point values, which improves performance proportionally with the bitwidth of IFMs based on bit-serial computing. But it only truncates the bitwidth while ignoring the bit sparsity of IFMs. By comparison, Bit-Balance achieves higher performance by skipping the zero bits of weights.\newline 
\indent\setlength{\parindent}{1em}4) Laconic\cite{8980351} tears MAC operation into IFMs and weights bit-serial computing and achieves great speedup by exploiting both IFM and weight sparsity. However, the distribution of zero bits in IFMs and weights can be irregular, which causes imbalanced workload in the PE array and degrades performance. By comparison, Bit-Balance achieves higher performance through bit-sparsity quantization.\\
\indent\setlength{\parindent}{1em}5) Bitlets\cite{10.1145/3466752.3480123} proposes a bit interleaving philosophy to maximally exploit bit-level sparsity, which enforces the  acceleration by decreasing the number of weights involved in computing. However, the logic corresponding to the bit-interleaving occupies 40\% area of the entire PE module. Besides, the computing units for the high-bit can be idle during low-bit operation, causing inefficient resource efficiency. By comparison, we achieved higher energy efficiency and resource efficiency with model-hardware co-design and adaptive bitwidth computing.\newline
\indent\setlength{\parindent}{1em}As for benchmarks, we conducted inference on AlexNet\cite{10.1145/3065386}, VGG-16\cite{simonyan2015deep}, ResNet-50\cite{7780459}, GoogleNet\cite{7298594}, Yolo-v3\cite{7780460} and compared our design with each baseline. The first four NNs are testified on ImageNet\cite{10.1145/3065386} and Yolo-v3 is testified on CoCo\cite{CoCo}.
\begin{table*}\centering
	\caption{Overall Comparison of Each Accelerator}
	\label{tab:tab5}
	\begin{tabular}{|c|c|c|c|c|c|c|c|}
		\hline
		Accelerator                                                      & \begin{tabular}[c]{@{}c@{}}Technology\\ (nm)\end{tabular} & \# of PEs & \begin{tabular}[c]{@{}c@{}}Area\\ (mm$^2$)\end{tabular}             & \begin{tabular}[c]{@{}c@{}}Frequency\\ (MHz)\end{tabular} & \begin{tabular}[c]{@{}c@{}}Peak Throughput\\ (GOP/s)\end{tabular}               & \begin{tabular}[c]{@{}c@{}}Power\\ (mW)\end{tabular}             & Classification                                                  \\ \hline
		\begin{tabular}[c]{@{}c@{}}Eyeriss\cite{7738524} \\ JSSCC-2017\end{tabular}    &  65nm                                                 & 168       & 12.25@full chip                                                  & 200                                                       & 33.6@16b MAC                                                                    & \begin{tabular}[c]{@{}c@{}}278@AlexNet\\ 234@VGG-16\end{tabular} & Bit Parallel                                                    \\ \hline
		\begin{tabular}[c]{@{}c@{}}Cambricon-X\cite{7783723}\\ MICRO-2016\end{tabular} &  65nm                                                 & 16        & 6.38@full chip                                                   & 1000                                                      & 256@16b MAC                                                                     & 278                                                              & \begin{tabular}[c]{@{}c@{}}Sparse \\ Bit Parallel\end{tabular} \\ \hline
		\begin{tabular}[c]{@{}c@{}}Stripe\cite{7529197}\\ CAL-2017\end{tabular}          &  65nm                                                 & 4096      & 122.1@full chip                                                  & 980                                                       & 64225@16b shift-add                                                             & -                                                                & Bit Serial                                                      \\ \hline
		\begin{tabular}[c]{@{}c@{}}Laconic\cite{8980351}\\ ISCA-2019\end{tabular}      &  65nm                                                 & 512       & 4.1@CC                                                           & 1000                                                      & 8192@1b shift-add                                                               & -                                                                & \begin{tabular}[c]{@{}c@{}}Sparse \\ Bit Serial\end{tabular}    \\ \hline
		\begin{tabular}[c]{@{}c@{}}Bitlet\cite{10.1145/3466752.3480123}\\ MICRO-2021\end{tabular}      &  65nm                                                 & 32        & 5.80@CC                                                          & 1000                                                      & 768@24b shift-add                                                               & \begin{tabular}[c]{@{}c@{}}1390@16b\\ 1199@ 8b\end{tabular}      & \begin{tabular}[c]{@{}c@{}}Sparse \\ Bit Serial\end{tabular}    \\ \hline
		Bit-balance                                                      &  65nm                                                 & 1024      & \begin{tabular}[c]{@{}c@{}}4.99@full chip\\ 2.91@CC\end{tabular} & 1000                                                      & \begin{tabular}[c]{@{}c@{}}1024@16b shift-add \\ 2048@8b shift-add\end{tabular} & \begin{tabular}[c]{@{}c@{}}820@16b \\ 857@ 8b\end{tabular}       & \begin{tabular}[c]{@{}c@{}}Sparse \\ Bit Serial\end{tabular}    \\ \hline
	\end{tabular}
\end{table*}

\subsection{Performance}
\begin{figure}[H]
	\centering
	\includegraphics[width=0.99\linewidth]{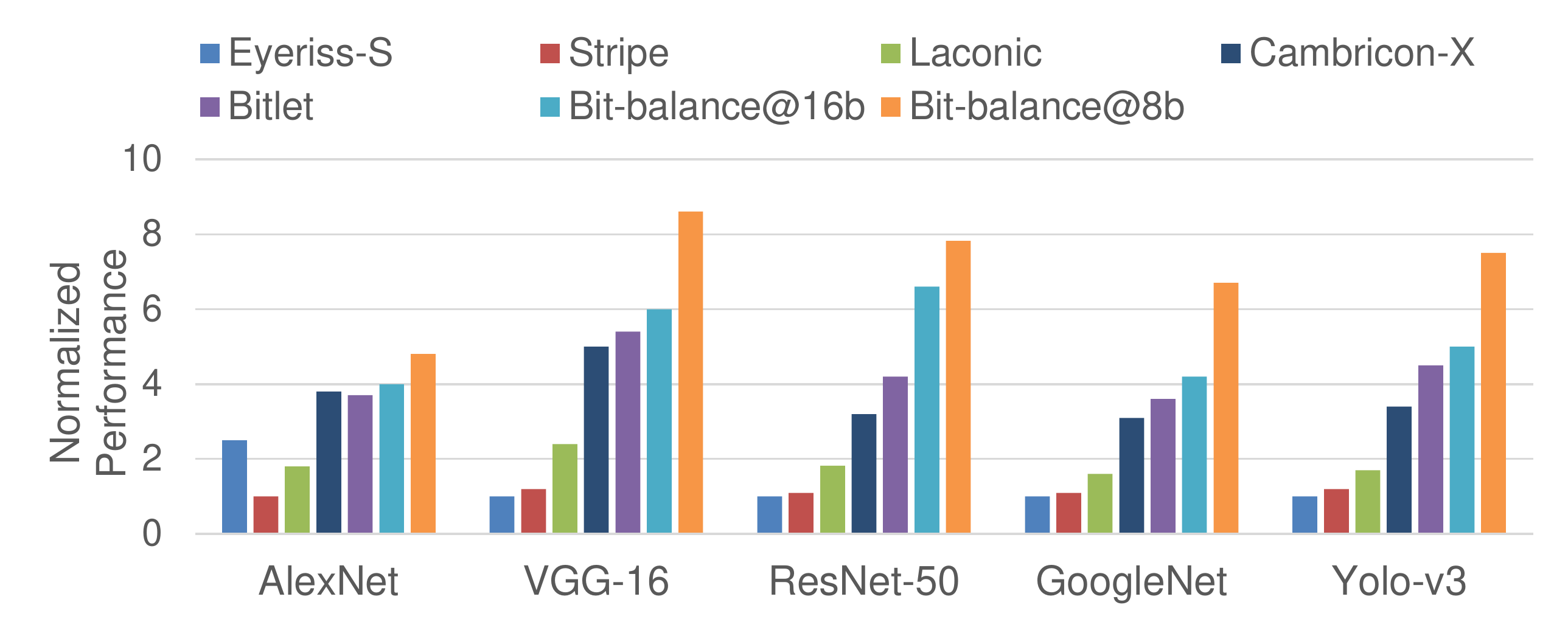}
	\caption{Performance comparison with Eyeriss, Cambricon-X, Stripe, Laconic, and Bitlets.}
	\label{fig10}
\end{figure}

To show the advantages of sparse processing methods in Bit-balance, we compared our design against Eyeriss\cite{7738524}, Cambricon-X\cite{7783723}, Stripe\cite{7529197}, Laconic\cite{8980351}, and Bitlets\cite{10.1145/3466752.3480123} on normalized performance, achieving $1.6\times$\textasciitilde$8.6\times$, $1.1\times$\textasciitilde$2.4\times$, $4\times$\textasciitilde$7.1\times$, $2.2\times$\textasciitilde$4.3\times$, and $1.1\times$\textasciitilde$1.9\times$ speedup, respectively, as shown in Fig.\ref{fig10}. In AlexNet, VGG-16, ResNet-50, GoogleNet, and Yolo-v3, we set the maximum NNZB, $N_{nzb\_max}$, as 3\textasciitilde4 in 16-bit precision and 4\textasciitilde5 in 8-bit precision, respectively, as shown in Tab.\ref{tab:tab6}. The columns of Top 1 and Top 5 accuracy show the inference accuracy of each NN and its corresponding accuracy loss in the brackets, which is referred to the training structures of Pytorch\cite{Pytorch}. The performance is calculated by the ratio of frequency and total cycles of inference computing, achieving $4\times$\textasciitilde$8\times$ speedup compared with the basic bit-serial computing of 16-bit precision. Though the accuracy of 16-bit precision is higher than the 8-bit and its maximum NNZB is also smaller, the peak throughput of the 8-bit is twofold, resulting in higher performance naturally. The normalized performance, $R_{p}$, is calculated by the ratio of our performance and other accelerators. Since Cambricon-X, Stripe, and Laconic didn't provide their performance in the paper, the normalized performance is calculated based on the comparison with Bitlet and Eyeriss in our benchmarks. \\
\begin{table*}\centering
	\caption{IFM and Weight Sparsity Ratios and Accuracy Loss of Each NN}
	\label{tab:tab6}
	\begin{tabular}{|c|cccc|cccc|}
		\hline
		\multirow{2}{*}{\begin{tabular}[c]{@{}c@{}}\\ NN Types\end{tabular}} & \multicolumn{4}{c|}{16-bit Precision}                                                                                                                                                                                                                                                    & \multicolumn{4}{c|}{8bit-precision}                                                                                                                                                                                                                                                        \\ \cline{2-9} 
		& \multicolumn{1}{c|}{$N_{nzb\_max}$} & \multicolumn{1}{c|}{\begin{tabular}[c]{@{}c@{}}Performance\\ (frame/s)\end{tabular}} & \multicolumn{1}{c|}{\begin{tabular}[c]{@{}c@{}}Top-1\\ Accuracy(\%)\end{tabular}} & \begin{tabular}[c]{@{}c@{}}Top-5\\ Accuracy(\%)\end{tabular} & \multicolumn{1}{c|}{$N_{nzb\_max}$} & \multicolumn{1}{c|}{\begin{tabular}[c]{@{}c@{}}Performance \\ (frame/s)\end{tabular}} & \multicolumn{1}{c|}{\begin{tabular}[c]{@{}c@{}}Top-1\\ Accuracy(\%)\end{tabular}} & \begin{tabular}[c]{@{}c@{}}Top-5\\ Accuracy(\%)\end{tabular} \\ \hline
		AlexNet                                                              & \multicolumn{1}{c|}{3}                  & \multicolumn{1}{c|}{270.5}                                                             & \multicolumn{1}{c|}{55.6(-0.9)}                                                  & 78.8(-0.3)                                                           & \multicolumn{1}{c|}{5}                  & \multicolumn{1}{c|}{326.2}                                                              & \multicolumn{1}{c|}{55.6(-0.9)}                                                  & 78.8(-0.3)                                                              \\ \hline
		VGG-16                                                               & \multicolumn{1}{c|}{3}                  & \multicolumn{1}{c|}{20.4}                                                              & \multicolumn{1}{c|}{70.8(-0.8)}                                                  & 89.9(-0.5)                                                           & \multicolumn{1}{c|}{4}                  & \multicolumn{1}{c|}{30.1}                                                               & \multicolumn{1}{c|}{71.2(-0.4)}                                                  & 90.1(-0.3)                                                              \\ \hline
		GoogleNet                                                            & \multicolumn{1}{c|}{4}                  & \multicolumn{1}{c|}{136.2}                                                             & \multicolumn{1}{c|}{69.1(-0.7)}                                                  & 89.3(-0.3)                                                           & \multicolumn{1}{c|}{5}                  & \multicolumn{1}{c|}{218.4}                                                              & \multicolumn{1}{c|}{69.1(-0.7)}                                                  & 89.3(-0.3)                                                              \\ \hline
		ResNet-50                                                            & \multicolumn{1}{c|}{3}                  & \multicolumn{1}{c|}{46.8}                                                              & \multicolumn{1}{c|}{75.2(-0.9)}                                                            &     92.4(-0.5)                                                                 & \multicolumn{1}{c|}{5}                  & \multicolumn{1}{c|}{56.3}                                                               & \multicolumn{1}{c|}{74.9(-1.2)}                                                            &     92.4(-0.5)                                                                \\ \hline
		Yolo-v3                                                            & \multicolumn{1}{c|}{3}                  & \multicolumn{1}{c|}{10.9}                                                              & \multicolumn{1}{c|}{61.9(-0.7)}                                                            &          -                                                            & \multicolumn{1}{c|}{4}                  & \multicolumn{1}{c|}{16.4}                                                               & \multicolumn{1}{c|}{60.3(-2.3)}                                                            &       -                                                               \\ \hline
	\end{tabular}
\end{table*}
\indent\setlength{\parindent}{1em}Compared with bit-parallel architectures, we leveraged bit-serial computing instead of 16-bit MAC unit. For Eyeriss, since it has been taped out, we assume the frequency of Eyeriss can reach to 1GHz for a fair comparison (Eyeriss-S in Fig.\ref{fig10}). Though we consumed multiple cycles for one MAC operation in a PE, our PE array size is larger than Eyeriss. Besides, the execution time of one MAC can be shrunk to $N_{nzb\_max}$ cycles by skipping zero bits. In the best case, we managed to achieve $30.1/(3.5) = 8.6\times$ speedup with 8-bit precision at VGG-16. For Cambricon-X, it exploits the weight element sparsity to accelerate computing. However, the sparsity of weights engaged in each PE can be imbalanced, causing performance degradation. Considering both sparsity exploiting and PE array size, we can achieve $1.1\times$\textasciitilde$2.4\times$ performance improvement. \\
\indent\setlength{\parindent}{1em}Compared with bit-serial architectures, we accelerate the NNs cooperated with bit-sparsity quantization. We only quantized the NNZB instead of the bitwidth in Strip. Therefore, our quantization method can accommodate wider numeric range of weight values and the NNZB in Bit-balance is smaller than the bitwidth in Stripe. Thus, we obtained $4\times$\textasciitilde$7.1\times$ speedup. Comparing with Laconic, we balanced the workloads across PE array by constraining the NNZB to a certain value. Though Laconic exploits sparsity of both IFMs and weights, its computing time is determined by the longest non-zero bit sequence, causing performance degradation once there exists a very long bit sequence involved in computing. Thus, we obtained $2.2\times$\textasciitilde$4.3\times$ performance improvement through load-balancing. For Bitlet, its performance improved by the bit-interleaving is similar with our method at the 16-bit precision. But we applied the adaptive datawidth computing for higher performance in low-bit operation, while in Bitlet, some computing units remain idle. In the best case, we obtained $56.3/29 = 1.9\times$ speedup in 8-bit precision at ResNet-50.

\subsection{Energy Efficiency and Resource Efficiency}
\begin{figure}[H]
	\centering
	\includegraphics[width=0.99\linewidth]{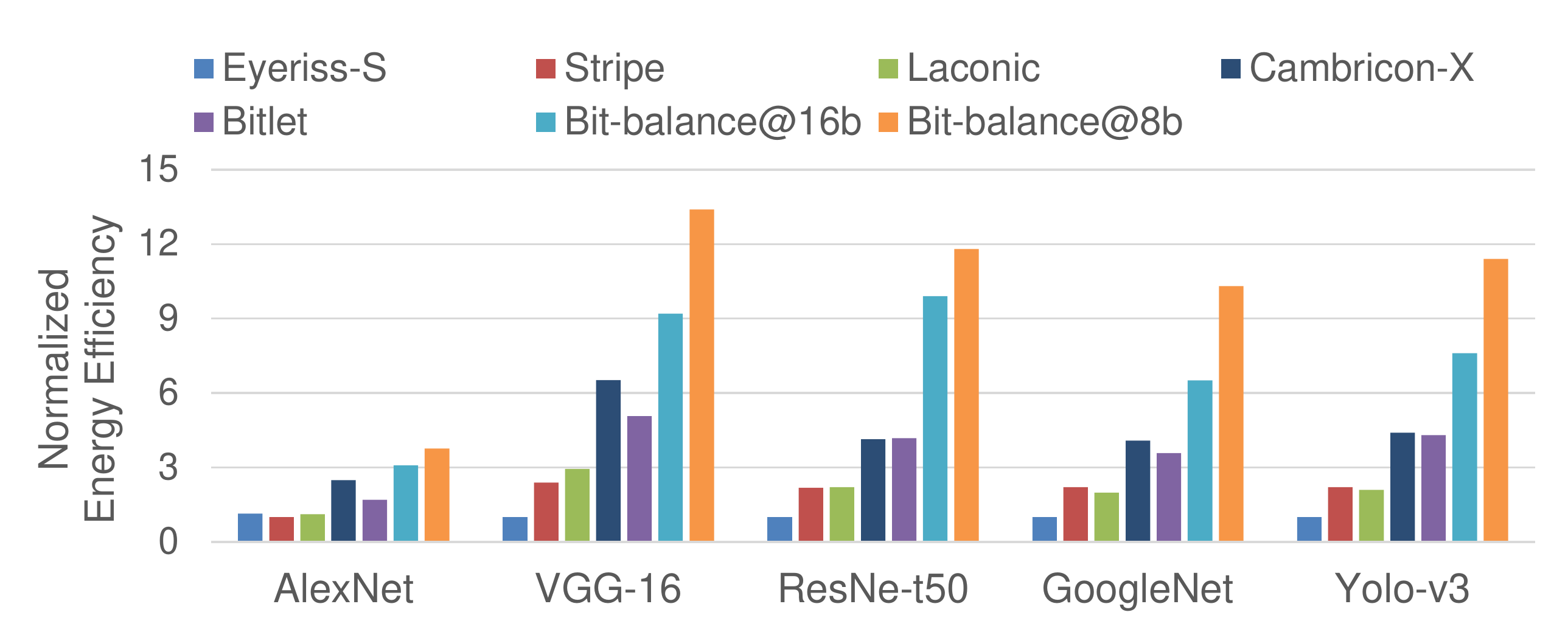}
	\caption{Energy efficiency comparison with Eyeriss, Cambricon-X, Stripe, Laconic, and Bitlets.}
	\label{fig11}
\end{figure}

\begin{figure}[H]
	\centering
	\includegraphics[width=0.99\linewidth]{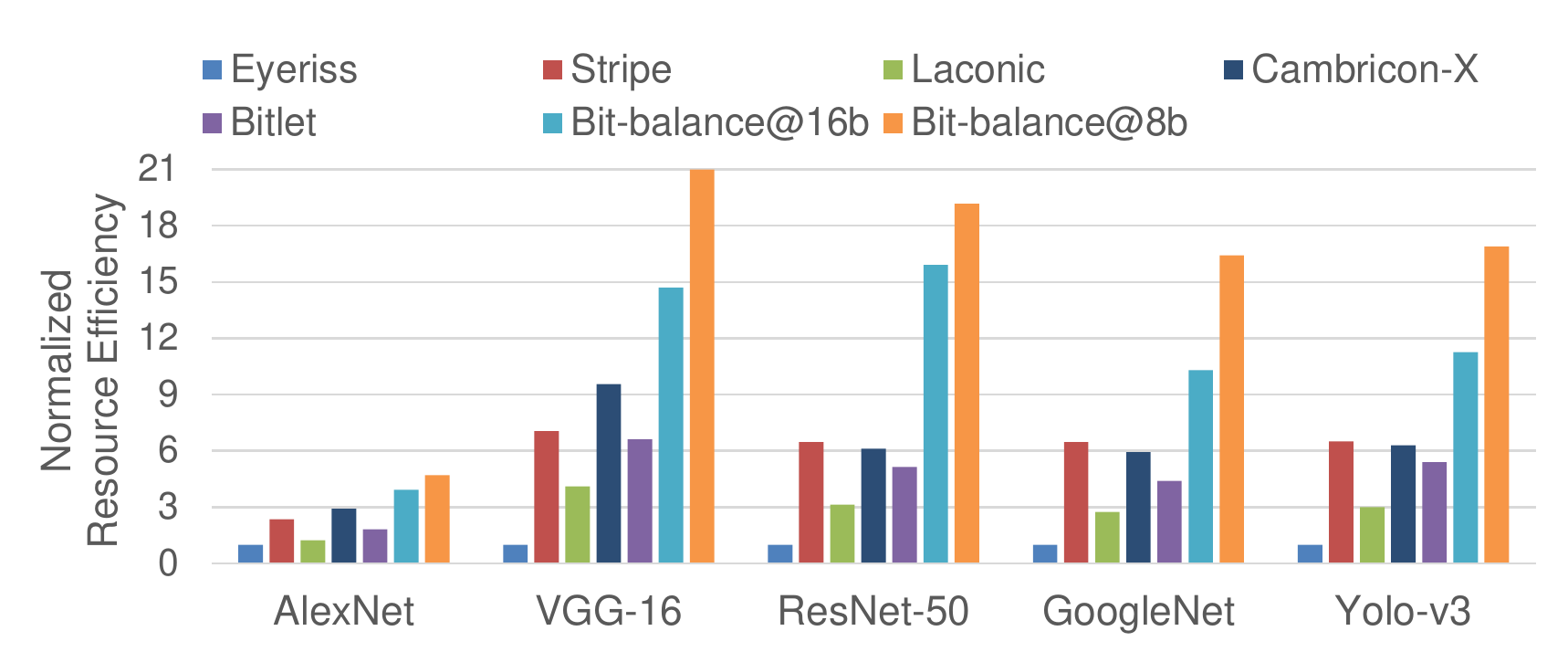}
	\caption{Resource efficiency comparison with Eyeriss, Cambricon-X, Stripe, Laconic, and Bitlets.}
	\label{fig12}
\end{figure}

Since the processing methods of each accelerator are different, it's essential to explore whether the benefits outperform the incurred overhead. We compared the energy efficiency and resource efficiency of Bit-balance with Eyeriss\cite{7738524}, Cambricon-X\cite{7783723}, Stripe\cite{7529197}, Laconic\cite{8980351}, and Bitlet\cite{10.1145/3466752.3480123}, achieving $2.7\times$\textasciitilde$11.3\times$, $1.3\times$\textasciitilde$2.8\times$, $3\times$\textasciitilde$5.6\times$, $2.7\times$\textasciitilde $5.4\times$, and $1.8\times$\textasciitilde $2.7\times$ energy efficiency improvement, respectively, as shown in Fig.\ref{fig11}, and achieving $3.9\times$\textasciitilde$21\times$, $1.6\times$\textasciitilde$3.9\times$, $1.7\times$\textasciitilde$3\times$, $3.2\times$\textasciitilde$6.3\times$, and $2.1\times$\textasciitilde $3.8\times$ resource efficiency improvement, respectively, as shown in Fig.\ref{fig12}. The energy efficiency is calculated as $R_{np}/R_{p}$, where $R_{np}$ and $R_{p}$ represent the ratio of Bit-balance and other baselines in normalized performance and power, respectively. For resource efficiency, the ratio of power, $R_{p}$, is replaced with the ratio of area, $R_{a}$, in the expression of energy efficiency. Since the PE array size of Stripe has been scaled for computing normalized performance, the expression, $R_{np}/R_{a}$ should multiply the ratio of peak performance, $R_{pp}$, representing the throughput per area.   \\
\indent\setlength{\parindent}{1em}Compared with fixed-point architectures, the MAC units consume more area and power than add-shift units in Bit-balance. For Eyeriss, the PE consists of a MAC unit and some memories, where the MAC unit only accounts for 9\% of total area and power, indicating that Eyeriss consumes much more resources and energy for one MAC operation than that in Bit-balance. Noting that we scale the frequency of Eyeriss to 1GHz, the power should be scaled up by $5\times$ while the area remains unchanged. In the best case, we achieve $8.6/(857/(236\times5)) = 13.4\times$ energy efficiency and $8.6/(4.99/12.25) = 21\times$ resource efficiency improvement in 8-bit precision at VGG-16. For Cambricon-X, it introduces the indexing module for sparse weights processing, accounting for 31\% of the total area and 34\% of the total power. However, the processing of sparse weights in Bit-balance are similar with dense bit-serial computing, inducing little overhead. Thus, we achieve $1.3\times$\textasciitilde$2.8\times$ energy efficiency and $1.6\times$\textasciitilde$3.9\times$ resource efficiency improvement.\\
\indent\setlength{\parindent}{1em}Compared with bit-serial architectures, the overhead of sparse computing in Bit-balance is smaller. For Stripe, it merged 16 shift-add units to one module, simplifying the intermediate steps, while the shift-add operations are performed individually in each PE of Bit-balance for sparse processing. Besides, since the PE array size scales by square while the memory scales linearly, the larger PE array size is, the less proportion memory accounts for. Thus, it consumes $(64225/1024)/(122.5/4.99) = 2.5\times$ less resource than Bit-balance for one add-shift operation. Combining the overhead with performance, we achieve $3\times$\textasciitilde$5.6\times$ energy efficiency and $1.7\times$\textasciitilde$3\times$ resource efficiency improvement. For Laconic, its PE unit decomposed the MAC into 16 pairs of IFM-bit and weight-bit operations. Though it can exploit the sparsity of both IFM and weight bit, many logic units are incurred for processing the intermediate results, resulting in $4.1/2.91 = 1.4\times$ computing core area overhead compared with Bit-balance. Therefore, we achieve $2.7\times$\textasciitilde $5.4\times$ energy efficiency and $3.2\times$\textasciitilde$6.3\times$ resource efficiency improvement. For Bitlet, its computing process consists of three steps: preprocessing, dynamic exponent matching and bit distillation. Each step is implemented in the corresponding modules, introducing 45\% area and 69\% power overhead. However, the preprocessing of weights in Bit-balance is executed offline, inducing no logic overhead. And the process of skipping zero bits is mastered by the top controller module, without inducing sparse processing logic into PEs. Thus, in the best case, we achieve $1.9/(857/1199) = 2.7\times$ energy efficiency and $1.9/(2.91/5.80) =3.8\times$ resource efficiency improvement in 8-bit precision at ResNet-50. \\

\begin{figure}[H]
	\centering
	\includegraphics[width=0.99\linewidth]{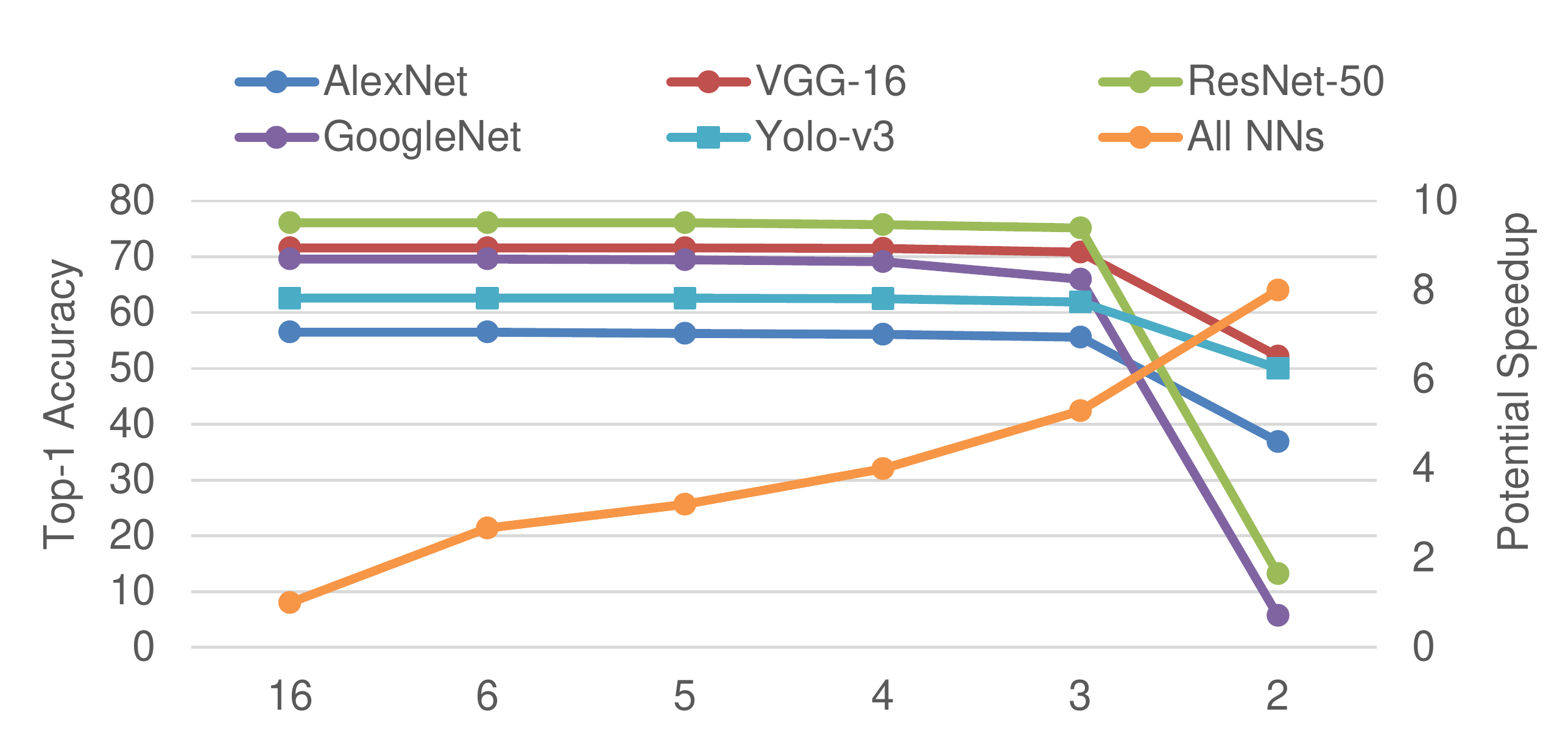}
	\caption{Potential Speedup and accuracy loss with different $N_{nzb\_max}$ in 16-bit precision.}
	\label{fig13}
\end{figure}

\begin{figure}[H]
	\centering
	\includegraphics[width=0.99\linewidth]{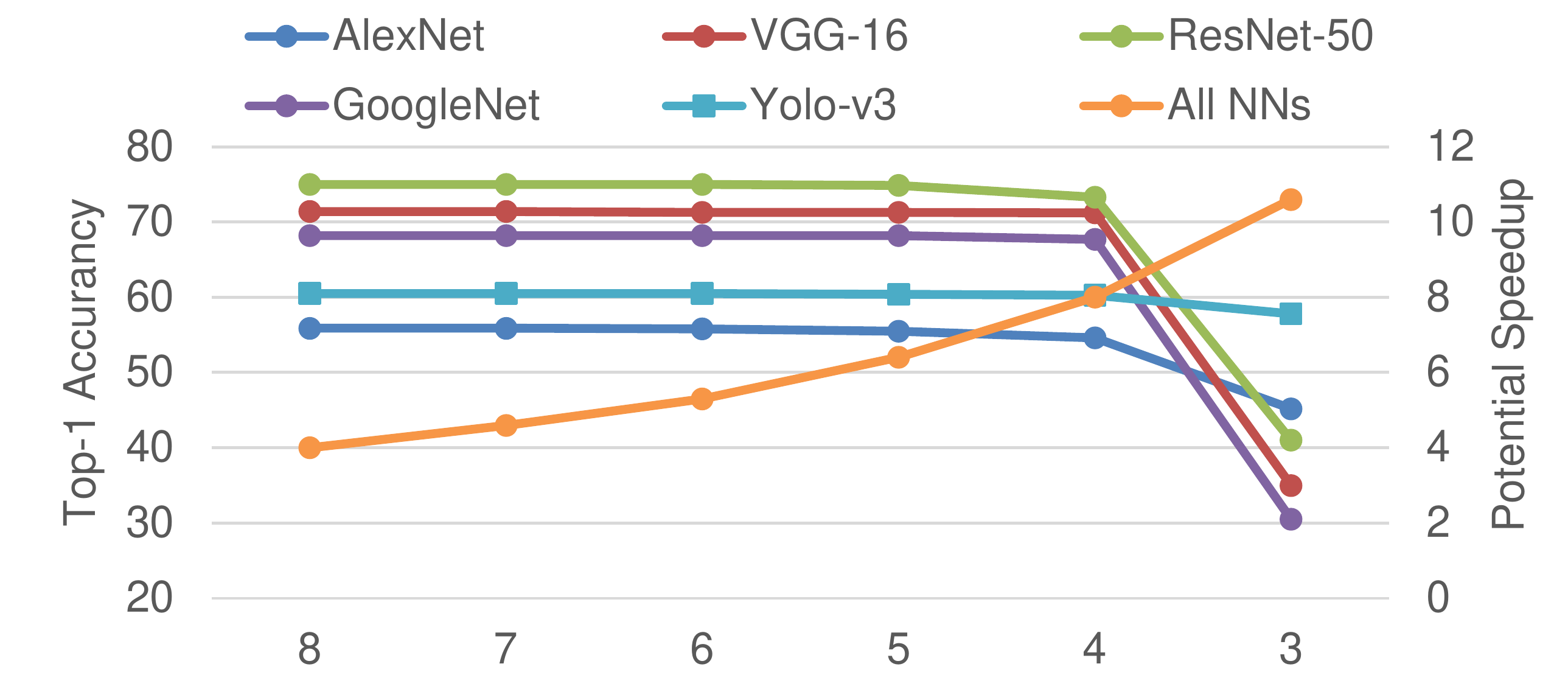}
	\caption{Potential speedup and accuracy loss with different $N_{nzb\_max}$ in 8-bit precision.}
	\label{fig14}
\end{figure}

\subsection{Sensitivity to Sparsity}

With higher sparsity of weight bits comes higher performance, under the sacrifice of accuracy. To explore the influence of weight-bit sparsity on performance and accuracy, we trained several groups of weights with different maximum NNZBs, as shown in Fig.\ref{fig13} and Fig.\ref{fig14}. The top-1 accuracy of float 32-bit precision in AlexNet, VGG-16, ResNet-50, GoogleNet, and Yolo-v3 are 56.5\%, 71.6\%, 76.1\%, 69.8\%, and 62.6\%, respectively referring to Pytorch\cite{Pytorch}. For 16-bit precision, the accuracy loss of each NN without bit-sparsity quantization is only 0\textasciitilde0.2\%. Then, we quantized the maximum NNZB to 2\textasciitilde6, achieving $2.67\times$\textasciitilde$8\times$ speedup. In the stage of $N_{nzb\_max}=6$\textasciitilde$4$, the accuracy of all NNs remains stable, with only 0.1\%\textasciitilde1.0\% loss. At the point of $N_{nzb\_max}=3$, the accuracy of GoogleNet drops about 3.7\%. When $N_{nzb\_max}=2$, the accuracy of all NNs precipitously drops. Thus, we chose $N_{nzb\_max}=3$ or $4$ for 16-bit precision in our work. \\
\indent\setlength{\parindent}{1em}The phenomenon in 8-bit precision is similar with the 16-bit precision. The 8-bit precision accuracy loss of each NN without bit-sparsity quantization is 0.2\%\textasciitilde2.1\%. In the stage of $N_{nzb\_max}=7$\textasciitilde$5$, the accuracy holds steady. At the point of $N_{nzb\_max}=4$, the accuracy of AlexNet, ResNet-50, and GoogleNet drops about 1.9\%\textasciitilde2.8\%. When $N_{nzb\_max}=3$, the accuracy of all NNs precipitously degrades. Thus, we chose $N_{nzb\_max}=4$ or $5$ for 8-bit precision computing.\\
\indent\setlength{\parindent}{1em}Though the performance improvement of 8-bit precision is higher than the 16-bit in Bit-balance owing to the adaptive bitwidth computing, there still exits some NNs of specific domains that are more suitable for 16-bit precision to maintain accuracy. For example, the accuracy of Yolo-v3 with 16-bit precision is 2\% higher than that of 8-bit precision on average, which is a significant accuracy improvement for NNs. Thus, adaptive bitwidth is essential for sparse bit-serial computing in NN acceleration.

\subsection{Comparison with Basic Bit-serial Architecture}
\begin{figure}[H]
	\centering
	\includegraphics[width=0.89\linewidth]{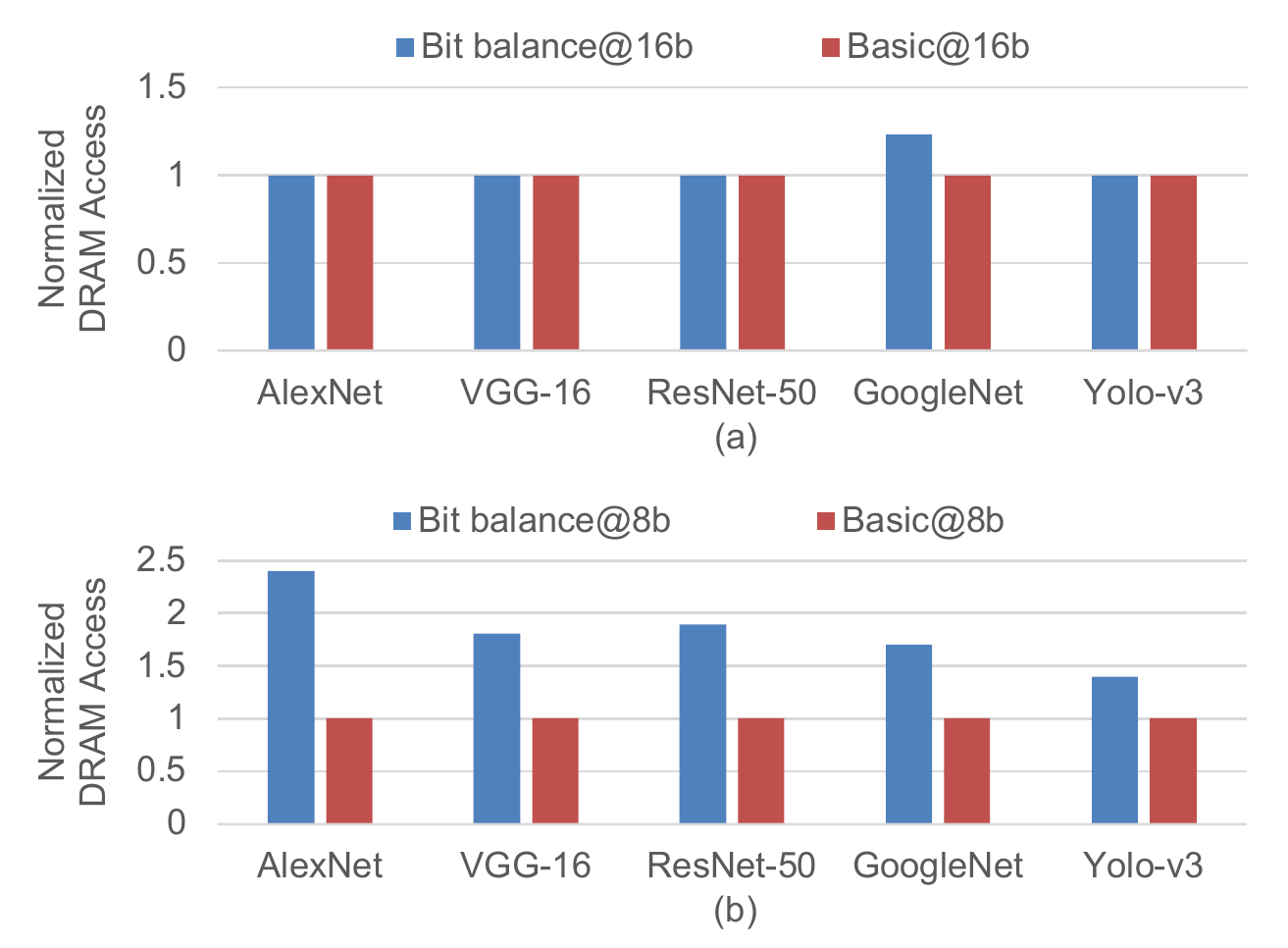}
	\caption{DRAM access comparison with basic bit-serial architecture. (a)16-bit precision. (b)8-bit precision.}
	\label{fig15}
\end{figure}

\begin{figure}[H]
	\centering
	\includegraphics[width=0.89\linewidth]{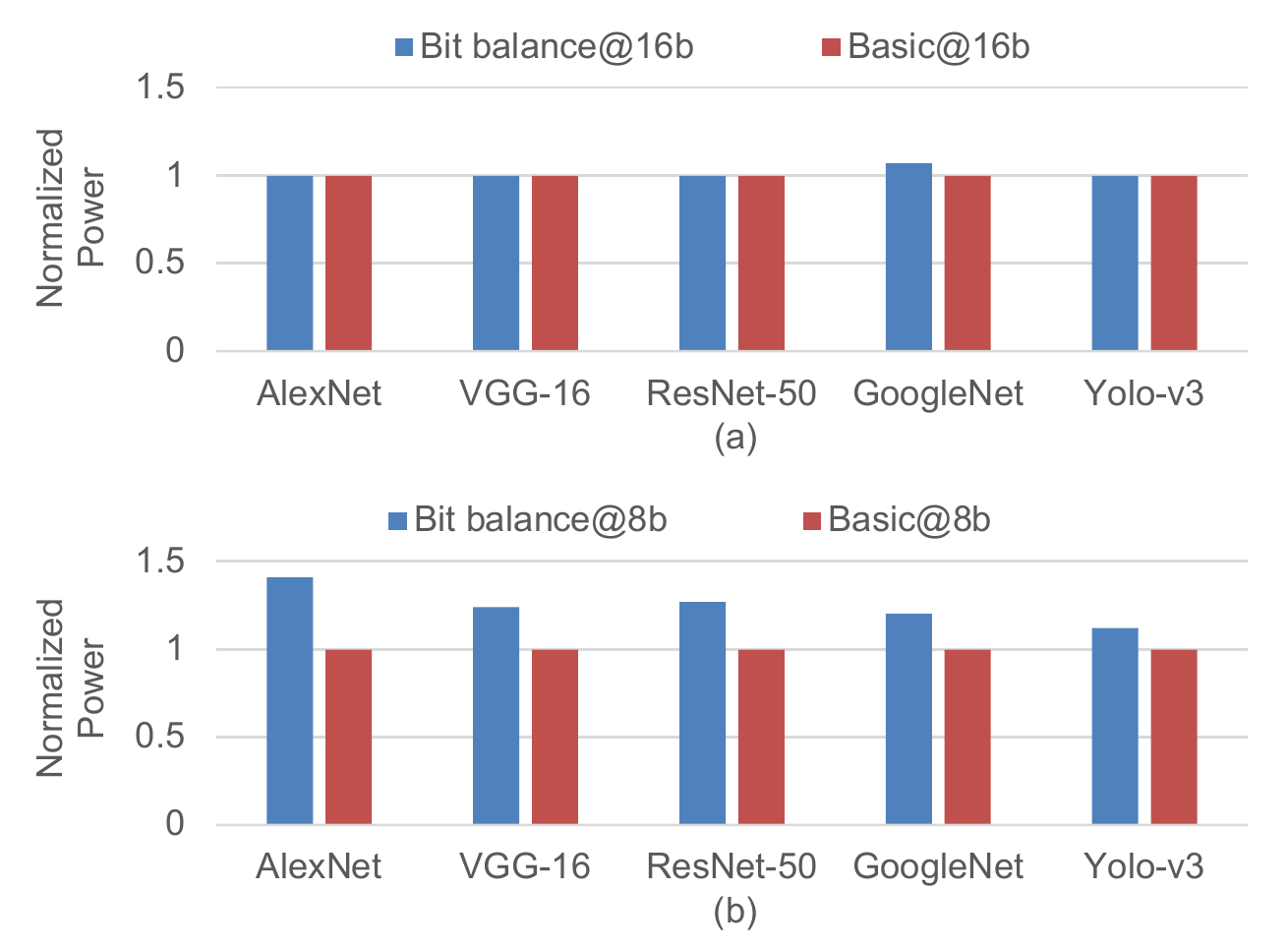}
	\caption{Power comparison with basic bit-serial architecture. (a)16-bit precision. (b)8-bit precision.}
	\label{fig16}
\end{figure}

\begin{figure}[H]
	\centering
	\includegraphics[width=0.89\linewidth]{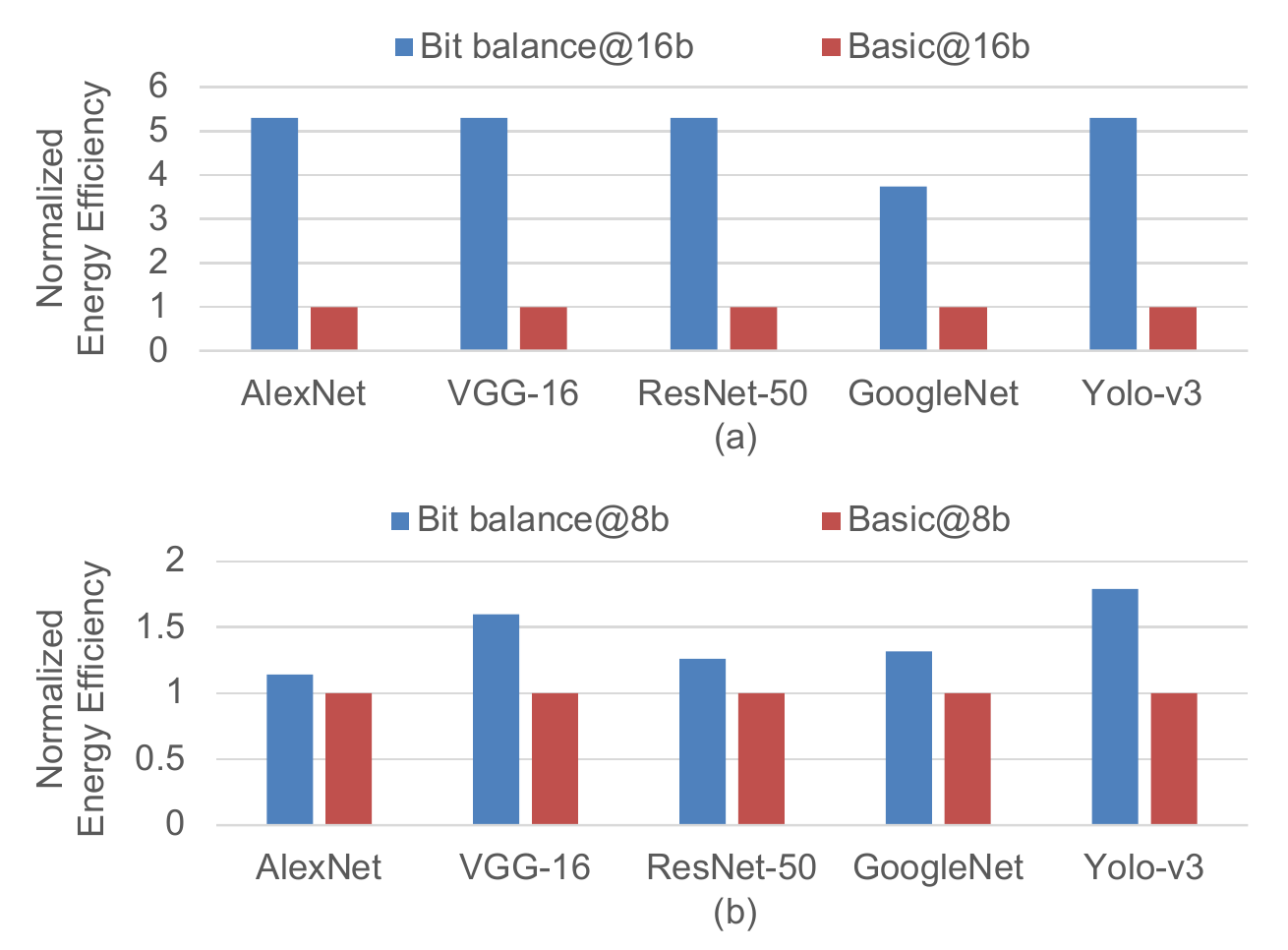}
	\caption{Energy efficiency comparison with basic bit-serial architecture. (a)16-bit precision. (b)8-bit precision.}
	\label{fig17}
\end{figure}

Since Bit-balance processes the weights with encoding format, it induces storage overhead and increases power consumption of DRAM access compared with original weights. To illustrate the benefit of sparse processing in Bit-balance, we compared the energy efficiency of the whole system with basic bit-serial architecture (Bit-balance without sparse processing). The power of DRAM access is estimated with CACTI\cite{10.1145/3085572} by the total DRAM access counts and runtime. Based on the dataflow in Section.V, though we consume $1\times$\textasciitilde$1.4\times$ power for $1\times$\textasciitilde$2.4\times$ DRAM access as shown in Fig.\ref{fig15} and Fig.\ref{fig16}, the energy efficiency of Bit-balance is still $1.14\times$\textasciitilde$5.3\times$ higher than basic bit-serial architecture owing to $1.6\times$\textasciitilde$5.3\times$ performance improvement as shown in Fig.\ref{fig17}. \\
\indent\setlength{\parindent}{1em}For 16-bit precision, the storage per weight of encoding format with $N_{nzb\_max} = 3$ is 16-bit, which allocates 1 bit for sign, 3 bits for bitmap, and $3\times4$ bits for weight-bit position. In the case of $N_{nzb\_max} = 4$, the storage per weight is $1+4+4\times4=21$-bit. Since the encoding format induces $1\times$\textasciitilde$1.3\times$ weight storage overhead compared with original 16-bit weight, the DRAM access increases $1\times$\textasciitilde$1.23\times$. Thus, Bit-balance consumes $1\times$\textasciitilde$1.07\times$ power. Owing to $4\times$\textasciitilde$5.3\times$ performance, the energy efficiency of Bit-balance is $3.73\times$\textasciitilde$5.3\times$ higher than basic bit-serial architecture. \\
\indent\setlength{\parindent}{1em}For 8-bit precision, the storage per weight with $N_{nzb\_max} = 4$ and $N_{nzb\_max} = 5$ are 17-bit and 21-bit respectively, inducing $2.1\times$\textasciitilde$2.6\times$ weight storage overhead compared with origianl 8-bit weight. Thus, Bit-balance consumes $1.12\times$\textasciitilde$1.41\times$ power for $1.4\times$\textasciitilde$2.4\times$ DRAM access. Owing to $1.6\times$\textasciitilde$2\times$ performance, the energy efficiency of Bit-balance is $1.14\times$\textasciitilde$1.79\times$ higher than basic bit-serial architecture.

\section{Conclusion}
This paper proposed a sparse bit-serial accelerator, called Bit-balance, for sparse weight bit processing, achieving significant performance improvement with low hardware cost. Meanwhile, we co-designed a bit-sparsity quantization method to maintain the maximum NNZB of weights to no more than a certain value with little accuracy loss, which can effectively balance the workloads of sparse weight bits in the PE array. Furthermore, to support adaptive bitwidth computing, we merged the 8-bit precision with 16-bit precision operation for higher resource efficiency. Compared with the previous sparse accelerators, Bit-balance can achieve better performance and energy efficiency with smaller area.

\bibliographystyle{IEEEtran}
\bibliography{bit_level}

\end{document}